\documentclass[
    ,final            
  ]
  {aipproc}

\layoutstyle{6x9}

\begin{document}

\title{The ${q}\overline{q}$ spectra and the structure
 of the scalar mesons}

\author{J. Vijande}{address={Grupo de F\' \i sica Nuclear,
Universidad de Salamanca, E-37008 Salamanca, Spain} }

\author{F. Fern\'{a}ndez}{
  address={Grupo de F\' \i sica Nuclear,
  Universidad de Salamanca, E-37008 Salamanca, Spain}
}

\author{ A. Valcarce}{
  address={Grupo de F\' \i sica Nuclear,
  Universidad de Salamanca, E-37008 Salamanca, Spain}
}

\begin{abstract}
The ${q}\overline{q}$ spectrum is studied within a
chiral constituent quark model. It provides with
a good fit of the available experimental data from
light (vector and pseudoscalar) to heavy mesons including some
recent results on charmonium. The $0^{++}$ light mesons and
the new $D$ states measured at different factories
cannot be described as $q\bar q$ pairs and a tetraquark
structure is suggested.  
\end{abstract}

\maketitle


\section{Introduction}

Since Gell-Man conjecture, most of the
meson experimental data were classified as $q\overline{q}$
states according to $SU(N)$ irreducible representations.
Nevertheless a number of interesting issues remains still open
as for example the understanding 
of the structure of the scalar mesons or the new $D_s$ states measured on
B factories.

The theoretical study of charmonium and bottomonium made clear that heavy-quark systems
are properly described by nonrelativistic potential models reflecting
the dynamics expected from QCD \cite{eich}.
The light meson sector has been studied by means of
constituent quark models, where quarks are dressed with a
phenomenological mass and bound in a nonrelativistic potential, usually
a harmonic oscillator \cite{isgu}.
Quite surprisingly a large number of properties of hadrons could be
reproduced in this way \cite{godf}. In this talk we present the
meson spectra obtained by means of a chiral constituent quark model
in a trial to interpret some of the still unclear experimental data
in the scalar sector. 
\section{SU(3) chiral constituent quark model}

Since the origin of the quark model hadrons have been considered to be
built by constituent (massive) quarks. Nowadays it is widely
recognized that because of the spontaneous
breaking of chiral symmetry in the light quark sector at
some momentum scale a  constituent quark mass $M(q^2)$
appears. Once a constituent quark mass is generated such particles have
to interact through $SU(3)$ Goldstone modes [pion, kaon, eta and sigma
(which simulates the two-pion exchange)]. Explicit expressions of these potentials can
be found elsewhere \cite{vij1}. 
In the heavy quark sector, chiral symmetry is
explicitly broken and therefore these interactions will not appear.
 
For higher momentum transfer quarks still interact through
gluon exchanges. Following de R\'{u}jula {\it et al.} \cite{ruju}
the one-gluon-exchange (OGE) interaction is taken as a
standard color Fermi-Breit potential.
In order to obtain a unified description of light, strange and
heavy mesons a scale dependent strong coupling constant has to be used \cite{bad}.
We parametrize this scale dependence by
 \begin{equation}
 \alpha_s(\mu)={\alpha_0\over{ln\left({{\mu^2+\mu^2_0}
 \over\Lambda_0^2}\right)}},
 \label{asf}
 \end{equation}
where $\mu$ is the reduced mass of the $q\bar q$ system and $\alpha_0$, $\mu_
0$ and $\Lambda_0$ are fitted parameters \cite{vij1}.
This equation gives rise to $\alpha_s\sim0.54$ for the light quark sector,
a value consistent with the one used in the study of the nonstrange
hadron phenomenology \cite{ente}, and
it also has an appropriate high $Q^2$ behavior \cite{pre0},
given a value of $\alpha_s\sim0.127$ at the $Z_0$ mass \cite{pre1}. The $\delta$ function
appearing in the OGE has to be regularized in order to avoid an unbound spectrum
from below. To solve numerically the Schr\"{o}dinger equation with this potential
we use a flavor-dependent regularization \cite{vij1}.

The other nonperturbative property of QCD is confinement.
Lattice QCD studies show that $q\overline{q}$ systems
are well reproduced at short distances by a linear potential that
is screened at large distances due to pair creation \cite{bali}.
One important question which has not been properly answered
is the covariance property of confinement. While the
spin-orbit splittings in heavy quark systems
suggest a scalar confining potential \cite{luca}, a significant mixture
of vector confinement has been used to explain the decay widths
of $P$-wave $D$ mesons \cite{suga}.
Such property, being irrelevant for the central part
of the interaction, determines the sign and strength of the
spin-orbit Thomas precession term which is important for the
scalar mesons. Therefore, we write
the confining interaction
as an arbitrary combination of scalar and vector terms
$V^{SO}_{CON}(\vec{r}_{ij})= (1-a_s)
V^{SO}_V(\vec{r}_{ij}) + a_s V^{SO}_S(\vec{r}_{ij})$
where $V^{SO}_{V}(\vec r)[V^{SO}_{S}(\vec r)]$ is the vector (scalar)
spin-orbit contribution.   

\section{Results}

With the quark-quark interaction described above we have solved
the Schr\"{o}dinger equation for the different $q\overline{q}$ systems.
Most of the parameters of the Goldstone boson fields are taken 
from the $NN$ sector. The eta
and kaon cutoff masses are related with the sigma and pion one as explained
in Ref. \cite{muli}: $\Lambda [u(d)s] \simeq \Lambda(ud)+m_{s}$,
$m_{s}$ being the strange quark current mass. 
The confinement parameters $a_{c}$ and $\mu_{c}$ are fitted to
reproduce the energy difference between the $\rho$ meson and its first
radial excitation and the $J/\psi$ and the $\psi (2S)$. The
parameters involved in the OGE are obtained from
a global fit to the hyperfine splittings well established in 
the Particle Data Group (PDG) \cite{pdgb}.
Finally, the relative strength of the scalar and
vector confinement is fitted to the energy of $a_{1}(1260)$ 
and $a_{2}(1320)$, ordering that cannot be reproduced with a pure
scalar confinement. We obtain $a_s=0.777$.

The spectra for the light pseudoscalar and
vector mesons and for heavy mesons have been reported in Ref. \cite{dub}. The agreement with experimental
data is remarkable. Let us emphasize that with only 11 parameters we are
able to describe more than 110 states. 

Recently Belle and BaBar \cite{bel1,bel2,bel3} collaborations have reported new experimental
measurements for the mass of the $\eta _{c}(2S)$. The average value from the
different measurement is significantly larger than most predictions of constituent
quark models and the previous experimental value of the PDG: 
$M[\eta_c(2S)]=3594 \pm 5$ MeV.
Such value cannot be easily explained
in the framework of constituent quark models because the resulting $2S$
hyperfine splitting (HFS) would be smaller than the predicted for the $1S$
states \cite{eber,hfs3}. Based on this fact some authors have claimed for
an $\alpha _{s}$ coupling constant depending on the radial excitation.

We predict a value $M[\eta _c(2S)]=$ 3627 MeV, within
the error bar of the last two Belle measurements, the ones
obtained with higher statistics. Moreover the ratio 2S to 1S HFS is
found to be 0.537, in good agreement with the experimental
data, 0.479. The reason for this agreement can be found in the shape of the 
confining potential that also influences the HFS, the
linear confinement being not enough flexible to
accommodate both excitations \cite{pedr}.

Other puzzling state is a narrow resonance around 2317 MeV, known as $D_{S_{J}}^{\ast }(2317)$ 
reported by BaBar \cite{baba}. This state has been confirmed by CLEO \cite{cleo} together with another
possible resonance around 2460 MeV. Both
experiments interpret these resonances as $J^P=0^+$ and $1^+$
states. This discovery has triggered a series of articles presenting alternative hypothesis \cite{arta}. 
The most striking aspect of these two resonances is that their masses are
much lower than expected. We obtain a mass of 
2470.9 MeV for the $J^P=0^+$ and 2565.5 MeV for the  $J^P=1^+$.
They are far from the experimental data 
although the rest of the states  ($1^+$,$2^+$,$1^-$ and $0^-$) 
agree reasonably well with the
values of the PDG for both for the $D$'s and $D_{s}$'s states.

Concerning the scalar sector our results are
shown in Table \ref{t1} (column three).
 We observe that for 
isovector states, there appears a candidate
for the $a_{0}(980)$, the $^3P_0$ member of the lowest $^3P_J$ 
isovector multiplet. The other candidate, the $a_{0}(1450)$, 
is predicted to be the scalar member of a $^3P_J$ excited isovector multiplet. 
This reinforces the predictions of the naive quark model, where 
the $LS$ force makes lighter the $J=0$ states with respect to the $J=2$.
The assignment of the $a_0(1450)$ 
as the scalar member of the lowest $^3P_J$ multiplet
would contradict this idea, because the $a_2(1312)$ is well established
as a $q \bar q$ pair. The same behavior is evident in the
$c\overline{c}$ and the $b\overline{b}$ spectra, making impossible to
describe the $a_{0}(1450)$ as a member of the lowest $^3P_J$
isovector multiplet without spoiling the description 
of heavy-quark multiplets.
However, in spite of the correct description of the mass 
of the $a_{0}(980)$, the model predicts a pure light-quark content,
what seems to contradict some experimental evidences. 
The $a_{0}(1450)$ is predicted to be also a pure light quark structure
obtaining a mass somewhat higher than the experiment.

In the case of the isoscalar states, one finds a candidate for the $f_{0}(600)$
with a mass of 402 MeV, in the lower limit of the
experimental error bar and with a strangeness content around 
$8\%$. The $f_{0}(980)$ and $f_{0}(1500)$ cannot be found for any
combination of the parameters of the model. 
It seems that a different structure rather
than a naive $q\overline{q}$ pair is needed to describe
these states. The $f_0(1500)$ is a clear
candidate for the lightest glueball \cite{amsl} and our results support this
assumption. For the $f_{0}(1370)$ state (which may actually correspond to
two different states \cite{blac}) we obtain two almost degenerate states
around this energy, the lower one with a predominantly nonstrange content,
and the other with a high $s\overline{s}$ content. 
Finally a state corresponding to the $f_{0}(1710)$ is obtained.

In the $I=1/2$ sector, as a consequence of the larger mass of the
strange quark as compared to the light ones, our model always predicts a mass
for the lowest $0^{++}$ state 200 MeV greater than the 
$a_{0}(980)$ mass. Therefore, being the $a_0(980)$ the member of the
lowest isovector scalar multiplet, the $\kappa (900)$
cannot be explained  as a $q\bar q$ pair.
We find a candidate for the $K_0^*(1430)$ although with a smaller mass.

\section{The scalar mesons as tetraquark}

Unlike the $q \bar q$ pairs tetraquark structures, 
suggested twenty years ago by Jaffe \cite{jaff}, 
can couple to $0^{++}$ without orbital excitation
and therefore could be serious 
candidates to explain the structure of the scalar mesons.

In this section we study tetraquark bound states by
solving the Schr\"odinger equation 
using a variational method where the spatial trial
wave function is a linear combination of gaussians. 
The technical details are given in Ref. \cite{vij2}. 
Due to the presence of the kaon-exchange
there is a mixture among different configurations
with the same isospin. In particular, in the isoscalar sector
the configurations: $[(qq)(\bar q \bar q)]$,
$[(qs)(\bar q \bar s)]$, and
$[(ss)(\bar s \bar s)]$ are mixed. The same happens in the isovector
case for the configurations:
$[(qq)(\bar q \bar q)]$, and
$[(qs)(\bar q \bar s)]$, and in the $I=1/2$ case for the configurations:
$[(qq)(\bar q \bar s)]$, and
$[(qs)(\bar s \bar s)]$. In all cases $q$ stands for a $u$ or $d$ quark.

The results are shown in Table \ref{t1} (column four) where we present the lowest
states for the three isospin sectors. As one can see, there appear two states, in the
isoscalar and isovector sectors, with almost the same mass, 
although too high to be identified with the $f_0(980)$
and $a_0(980)$. In the $I=1/2$ sector, there appears a
candidate to be identified with the $\kappa(900)$. 
It has been recently argued the possible importance of 
three-body forces arising from the confining interaction 
for those systems containing at least three quarks \cite{dimi}. We have
performed a calculation including a three-body confining
term as the one reported in Ref. \cite{dimi} fixing its
strength to reproduce the mass of the $f_0(980)$. The results are shown in
Table \ref{t1} (column five) and as can be seen the
degeneracy between the isoscalar and isovector states remains although their
masses are now compatible with the experimental data.
The lowest state of the isoscalar and $I=1/2$ sectors are almost not affected.

\begin{table}[ht]
\caption{Light scalar meson masses in MeV}
\label{t1}
\begin{tabular}{cccccc}
\hline
$(q\bar q)$ state $(n^{2I+1,2S+1}L_J)$ & Meson & $(q\bar q)$ & $(4q)$ &$(4q)$+Three body & Experiment \\ 
\hline
$1^{3,3}P_0$ & $a_0(980)$ & 983.5 &1343&968& 984.7$\pm$1.2 \\ 
$2^{3,3}P_0$ & $a_0(1450)$ & 1586.3 &&& 1474$\pm$19 \\ 
$1^{1,3}P_0$ & $f_0(600)$ & 402.7 &604&644& 400$-$1200 \\ 
& $f_0(980)$ & &1325&1007&980$\pm$10 \\ 
$1^{1,3}P_0$ & & 1341.7 &&&\\ 
$2^{1,3}P_0$ & $f_0(1370)$ & 1391.2 &&& 1200$-$1500 \\ 
$2^{1,3}P_0$ & $f_0(1710)$ & 1751.8 &&& 1713$\pm$6 \\ 
$3^{1,3}P_0$ & $f_0(2020)$ & 1893.8 &&& 1992$\pm$16 \\ 
& $\kappa(900)$ & &1026&922& 797$\pm$43\\ 
$1^{2,3}P_0$ & $K^*_0(1430)$ & 1213.5 &&& 1412$\pm$6 \\ 
$2^{2,3}P_0$ & $K^*_0(1950)$ & 1768.5 &&& 1945$\pm$30 \\ \hline
\end{tabular}
\end{table}

Using the same interaction and formalism we have calculated
the $D_{S_{J}}^{\ast }(2317)$ as a $[(uc)(\bar u \bar s)]$
tetraquark. The result we obtain, M=2389 MeV, suggests that 
this state could also have a significant tetraquark component.

As a summary, we have found tetraquark bound states in the region
of the light scalar mesons and in the $D_{S_{J}}^{\ast }(2317)$. Our results suggest
that some states, as it is the case of the 
$a_0(980)$ and $f_0(600)$, could present a significant mixture of
$q\bar q$ and tetraquark structures, but it assigns a clear tetraquark structure to the 
$f_0(980)$ and the $\kappa(900)$. 
However, more accurate
calculations including the exchange terms in the variational wave function
which are negligible for the heavy-light tetraquarks
and the explicit coupling to $q \bar q$ channels should be done 
before drawing any definitive conclusion.


\begin{theacknowledgments}
This work has been partially funded by Ministerio de 
Ciencia y Tecnolog{\'{\i}}a under Contract No. BFM2001-3563
and by Junta de Castilla y Le\'{o}n under
Contract No. SA-109/01.
\end{theacknowledgments}


\bibliographystyle{aipprocl} 


\IfFileExists{\jobname.bbl}{}
 {\typeout{}
  \typeout{******************************************}
  \typeout{** Please run "bibtex \jobname" to optain}
  \typeout{** the bibliography and then re-run LaTeX}
  \typeout{** twice to fix the references!}
  \typeout{******************************************}
  \typeout{}
 }

\end{document}